\documentclass[twocolumn,showpacs,preprintnumbers,amsmath,amssymb]{revtex4}

\usepackage{graphicx}
\usepackage{dcolumn}
\usepackage{bm}

\begin{document}


\title{Excitable Scale Free Networks}

\author{Mauro Copelli}
 \email{mcopelli@df.ufpe.br}
\affiliation{
Laborat\'orio de F\'{\i}sica Te\'orica e Computacional, Universidade Federal de Pernambuco, Departamento
 de F\'{\i}sica, 50670-901 Recife, PE, Brazil
}%

\author{Paulo R. A. Campos}
 \email{paulo.campos@df.ufrpe.br}
\affiliation{%
Departamento de F\'{\i}sica, Universidade Federal Rural de Pernambuco, 
52171-900 Recife, PE, Brazil
}%

\date{\today}

\begin{abstract}
When a simple excitable system is continuously stimulated by a
Poissonian external source, the response function (mean activity
versus stimulus rate) generally shows a linear saturating shape. This
is experimentally verified in some classes of sensory neurons, which
accordingly present a small dynamic range (defined as the interval of
stimulus intensity which can be appropriately coded by the mean
activity of the excitable element), usually about one or two decades
only. The brain, on the other hand, can handle a significantly broader
range of stimulus intensity, and a collective phenomenon involving the
interaction among excitable neurons has been suggested to account for
the enhancement of the dynamic range. Since the role of the pattern of
such interactions is still unclear, here we investigate the
performance of a scale-free (SF) network topology in this dynamic
range problem. Specifically, we study the transfer function of
disordered SF networks of excitable Greenberg-Hastings cellular
automata.  We observe that the dynamic range is maximum when the
coupling among the elements is critical, corroborating a general
reasoning recently proposed. Although the maximum dynamic range
yielded by general SF networks is slightly worse than that of random
networks, for special SF networks which lack loops the enhancement of
the dynamic range can be dramatic, reaching nearly five decades. In
order to understand the role of loops on the transfer function we
propose a simple model in which the density of loops in the network
can be gradually increased, and show that this is accompanied by a
gradual decrease of dynamic range. 
\end{abstract}

\maketitle

\section{\label{intro}Introduction}

Recent applications of tools from Statistical Physics have brought
about new perspectives to theoretical Neuroscience. On the one hand,
networks of simplified neuron models seem to capture essential
features of collective neuronal
dynamics~\cite{lewis00,Copelli02,Copelli05a,Copelli05b,Haldeman05},
very often being also amenable to analytical calculations via mean
field approximations or Fokker-Planck
equations~\cite{Furtado06,Kinouchi06a,Doiron06}. On the other hand,
experimental data from real neural networks have yielded extremely
interesting results when analyzed within the framework of complex
networks, often revealing the small-world character for structural
(i.e. anatomical) connectivity in different spatial
scales~\cite{Watts98,Amaral00,Sakata05}, as well as scale-free
characteristics for functional connectivity both in {\it
in-vitro}~\cite{Beggs03} and in fMRI~\cite{Eguiluz05, Chialvo04} data
(see~\cite{Sporns04a} for a recent review).

In the context of modelling, one intriguing question in Neuroscience
regards the stunning ability that brains have to cope with sensory
stimuli that vary over many orders of
magnitude~\cite{Kinouchi06a}. The experimental evidence supporting
this claim has been accumulating for about a century in the
Psychophysics literature: the perception of a given stimulus grows
with a power law of the stimulus intensity (Stevens law), with an
exponent (Stevens exponent) which is typically $<1$, implying
low-stimulus amplification and large dynamic range~\cite{Stevens}.
This is in stark contrast with the poor performance of single neurons:
as a function of the stimulus intensity, the mean firing rate of
sensory neurons experimentally shows the linear saturating shape that
one expects for general excitable systems, so their responses
consistently have a small dynamic range, typically about one or two
decades only~\cite{Firestein93, Rospars00, Rospars03, Deans02}. How
can these two experimental results be reconciled? 
A solution which has been proposed for this apparent paradox involves
a collective phenomenon. The idea is that if excitable elements with
small dynamic range are coupled, signal propagation in the network
amplifies the average activity, as compared to that of an isolated
node. This collectively leads to a significant enhancement of dynamic
range, thus providing a possible solution to a problem faced by
biological as well as artificial sensors: how to code for several
orders of magnitude of stimulus intensity, starting from narrow-coding
elements~\cite{Copelli02, Copelli05a, Copelli05b, Furtado06,
Kinouchi06a}.


The reasoning underlying the enhancement of dynamic range is very
general and applies to essentially any network topology. Consider the
limit of very weak stimulus, where each excitable element has a small
probability of being excited. By coupling the elements, a single
stimulus event will be amplified to neighboring sites, which will
further amplify it, and so forth. If the coupling strength is small,
this excitable wave will eventually die out, but the overall network
activity (the response to the stimulus) will still be larger than that
of the isolated element that originally received the stimulus. The
larger the coupling strength, the larger the amplification, and so the
sensitivity and the dynamic range of the response curve initially
increase with coupling. There is, however, a critical value of the
coupling above which self-sustained activity becomes stable. Above
this (typically second order) nonequilibrium phase transition, the
response of the network for weak stimulus is hindered, because it is
masked by the self-sustained activity of the network. This gets worse
and worse as the coupling increases, so above criticality the dynamic
range of the response curve decreases with increasing coupling
strength.  Therefore, the dynamic range is optimal at
criticality~\cite{Kinouchi06a}.

The above mechanism has been tested in regular~\cite{Copelli02,
Copelli05a, Copelli05b, Furtado06} as well as
random~\cite{Kinouchi06a} networks of excitable elements. The maximum
enhancement in dynamic range is about 100\% in one-dimensional
networks~\cite{Furtado06} and 50\% in random
graphs~\cite{Kinouchi06a}. It is not {\it a priori\/} clear how the
performance depends on the network topology and, in particular, which
one gives the best results. Given the potential applications of the
mechanism to artificial sensors, as well as the relevance to
Neuroscience, in this paper we study the performance of a scale-free
topology in the dynamic range problem. We show that a particular class
of scale-free networks, those with no loops, yield the best
performance known so far. We investigate the role of loops on the
dynamic range by introducing a slightly modified version of the
Barab\'asi-Albert scale-free model where we can now tune the amount of
loops in the network.


The paper is organized as follows. In section~\ref{model} we introduce
the model and give a precise definition of the dynamic range. Results
are discussed in section~\ref{results} for the standard
Barab\'asi-Albert scale-free model (\ref{ba}) as well as for a
slightly modified ``loop-diluted'' version that we introduce
(\ref{ld}). Our concluding remarks are presented in
section~\ref{conclusions}.




\section{\label{model}The model}


We consider a variant of the Greenberg-Hastings cellular
automaton~\cite{Greenberg78}, which is one of the simplest models of
an excitable system and can be used in large-scale simulations. In the
model, each excitable node $i=1,\ldots,N$ could represent either a
neuron, an active dendritic patch or even sub-cellular excitable
processes. Each node can be in one of $n$ states: $x_i=0$ is the
quiescent state (e.g. a polarized neuron), $x_i=1$ is the excited
state (e.g. a spiking neuron) and $x_i=2,\ldots,n-1$ are refractory
states (e.g. a hyperpolarized neuron). Once a site is excited
($x_i=1$), it deterministically goes through the next $n-2$ refractory
states, after which it jumps to the quiescent state $x_i=0$ (the
automaton is therefore cyclic~\cite{Marro99}). Each node is
independently excited by a stochastic external source, which mimics
the effect of an stimulus on the lattice. We model the arrival of a
suprathreshold stimulus by a Poisson process with rate $r$: at each
time step $\tau$ an attempt to stimulate a site occurs with
probability

\begin{equation}
\lambda = 1-\exp(-r\tau)
\end{equation}
(we adopt the arbitrary time scale of $\tau=1$~ms, which is the
characteristic time scale of a neuronal spike). We refer to the rate
$r$ as the stimulus intensity. In order to become excited in time
$t+\tau$ a given site has to be in state $0$ at time $t$. There are
two different ways by which a site can be excited: by the continuous
stimulation of the external source (with probability $\lambda$ per
time step) or by stimulus propagation from its excited neighbors. Thus
the probability that a quiescent site $i$ is excited in the next time
step is

\begin{equation}
P_i(t+\tau) = 1 - (1-\lambda)\prod_{j=1}^{k_i}(1-p_{ij})\delta(x_j(t),1)\; ,
\end{equation}
where $\delta(a,b)$ is the Kronecker delta, $k_i$ is the number of
neighbors (connectivity or degree) of site $i$ and $p_{ij}$ is the
probability that excitation from site $j$ gets transmitted to site
$i$. There is quenched disorder in the coupling: the probabilities
$p_{ij}$ are initially drawn from a uniform distribution in
$[0,2\sigma/K]$ if $2\sigma/K<1$, or $[2\sigma/K-1,1]$ if
$2\sigma/K>1$, where $K = \left<k\right>$ is the mean connectivity of
the network and $\sigma$ is the coupling parameter (for simplicity, we
consider the case of bidirectional coupling $p_{ij}=p_{ji}$). Note
that, in a mean field approximation, $\sigma$ coincides with the
branching ratio, defined as the average of the number of descendant
excitations divided by the number of ancestor excitations of each
site. Such mean field approximation provides a quite satisfactory
agreement with simulation results for random graph topologies (as
expected) and accurately predicts a phase transition at
$\sigma_c=1$~\cite{Kinouchi06a,Haldeman05}.

The mean firing rate of the network is defined as $F\equiv
T^{-1}\sum_t^T \rho_t$, where $\rho_t \equiv N^{-1}\sum_{i}^N
\delta(x_i(t),1)$ is the instantaneous density of active (excited)
sites and $T$ is a given time window for measurements (we have used
$N=10^4$, $T=10^4$ steps and $n=5$ states in most simulations). We
refer to $F(r)$ as the {\em response function\/} (or transfer
function) of the network. It typically shows the sigmoidal shape in a
log-linear scale exemplified in Fig.~\ref{fig:response}(a), with a
baseline activity $F_0 \equiv \lim_{r\to 0} F(r)$ and saturation at
$F_{max}\equiv \lim_{r\to\infty} F(r)$. The dynamic range $\Delta$ of
the response function is defined as the width (measured in dB) in
stimulus intensity $r$ which can be ``appropriately coded'' by $F$. In
the biological literature, this is usually operationalized as
follows~\cite{Firestein93,Rospars00}: by letting $F_x \equiv F_0 +
x(F_{max} - F_0)$, where $0 \leq x \leq 1$,
and $r_x$ be the corresponding stimulus intensity, ($F(r_x)=F_x$, see
triangles in Fig.~\ref{fig:response}(a) for an example), the dynamic
range is defined as
\begin{equation}
\Delta = 10\log_{10}\left(\frac{r_{0.9}}{r_{0.1}} \right)\; ,
\end{equation}
therefore excluding stimuli whose response is just above baseline
($r<r_{0.1}$) or too close to saturation ($r>r_{0.9}$). For an
isolated Greenberg-Hastings excitable node, one can easily show that
the dynamic range is $\Delta \lesssim
19$~dB~\cite{Copelli05b,Furtado06}.

\section{\label{results}Results}

\subsection{\label{ba}Barab\'asi-Albert networks}

We consider scale-free networks~\cite{Barabasi99Sci} of such excitable
elements. Several investigations show that distinct systems such as
World-Wide Web~\cite{Barabasi99Sci}, scientific~\cite{Newman01e},
metapopulation dynamics~\cite{Moreno02,Vuorinen04} and biochemical
networks~\cite{Jeong00metabolic,Jeong01lethality} self-organize into a
scale-free configuration \cite{Albert02}, which means that the
probability $P_{k}$ that a given node has $k$ edges follows a
power-law distribution like
\begin{equation}\label{scale}
P_{k} \propto k^{-\gamma}.
\end{equation}
Measurements in real systems estimate $\gamma$ in the range $[2,3]$.
Eq. (\ref{scale}) basically means that poorly-connected nodes are most
frequent in the network than well-connected nodes (hubs).

To establish scale-free networks, we follow the standard algorithm by
Barab\'asi and Albert (BA)~\cite{Barabasi99Sci}, which regards
preferential attachment and growth as mechanisms for the emergence of
the scale-free character. In this algorithm, the resulting networks
display connectivity distribution according to $P_{k} \propto k^{-3}$.
The parameters of the BA model are the number of nodes $N$ and $m$,
which corresponds to the number of links that a newly introduced node
adds to the network. These $m$ links are most probably attached to
those nodes with an already large number of edges.

\begin{figure}
\centerline{\includegraphics[width=0.95\columnwidth]{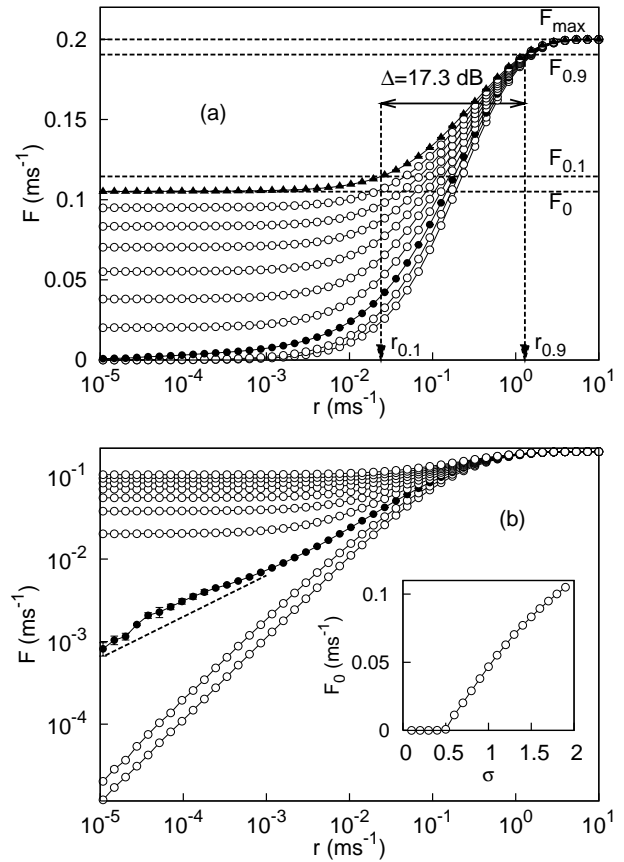}}
\caption{\label{fig:response}(a) Response functions for BA scale-free
networks with $m=10$: mean firing rate $F$ versus stimulus rate
$r$. Different curves denote different values of the branching
parameter $\sigma$: from bottom to top, $\sigma=0.1,
0.3,\ldots,1.9$. Filled circles: $\sigma=0.5$, which is close to the
critical value. The horizontal lines exemplify how the dynamic range
$\Delta$ is calculated for $\sigma=1.9$ (filled triangles). (b)
Log-log version of (a). The dashed line shows an exponent
$1/2$. Inset: Self-sustained activity $F_0$ versus $\sigma$,
illustrating the phase transition close to $\sigma_c\simeq 0.5$.}
\end{figure}

\begin{figure}
\centerline{\includegraphics[width=0.95\columnwidth]{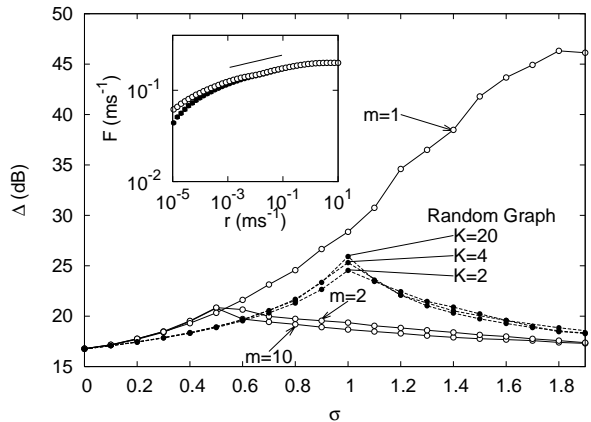}}
\caption{\label{fig:deltasigma}Dynamic range $\Delta$ versus branching
parameter $\sigma$ for distinct network topologies: BA scale-free
networks (open circles) and Erd{\H o}s-R\'enyi random graphs (filled
circles) with approximately the same mean connectivity. Inset:
Response function for BA scale-free networks with $m=1$ and $\sigma=2$
for $N=10^4$ (filled circles) and $N=3\times 10^4$ (open
circles). The solid line shows a slope $\simeq 0.07$.}
\end{figure}

Figure~\ref{fig:response} shows the results for $m=10$. For small
values of $\sigma$, the response function $F(r)$ increases linearly
for weak stimulus. This linearity can be easily interpreted: each
stimulus arrival generates an excitable wave that will have a finite
lifetime and will die before another wave is created. For stronger
stimulus (larger $r$) linearity breaks down, since there is
interaction among waves, which partially annihilate each other. For
very large $r$, the firing rates reach a saturation value which scales
with the inverse of the refractory period, $F_{max} =
1/n$~\cite{Copelli02, Copelli05a, Copelli05b}. As the value of
$\sigma$ increases, so does the lifetime of an excitable wave, leading
to larger amplification of weak stimuli and a corresponding
enhancement of dynamic range (see Fig.~\ref{fig:deltasigma} for
$\sigma \lesssim 0.5$). When $\sigma=\sigma_c$, the lifetime of the
excitable waves effectively diverges and the system undergoes a second
order phase transition (notice the change in the exponent in the
filled circles of Fig.~\ref{fig:response}(b)). For $\sigma>\sigma_c$,
any perturbation in the network will lead to a stable self sustained
activity, $F_0>0$ (inset of Fig.~\ref{fig:response}(b)) which, as
explained in section~\ref{intro}, leads to smaller values of the
dynamic range as the coupling increases~\cite{Kinouchi06a} (see
Fig.~\ref{fig:deltasigma} for $\sigma \gtrsim 0.5$).

One observes that, differently from random graph
topologies~\cite{Kinouchi06a}, the transition for scale-free excitable
networks occurs at $\sigma_c< 1$. We speculate that this is due to the
hubs, which have a local branching ratio $\sigma_i = \sum_j^{k_i}
p_{ij}$ larger than unit even for $\sigma<1$ and could therefore
facilitate the transition. It is also interesting to note that
deviations from mean field behavior have been predicted for the
contact process (CP) in a scale-free
network~\cite{Castellano06}. Apart from the refractory period and the
disorder, the CP is similar to the model we study here (in the sense
that it has a unique absorbing state with no symmetries). In
Fig.~\ref{fig:response}, however, the response exponent at criticality
(defined by $F(r;\sigma_c) \sim r^{1/\delta_{h}}$) seems to be
compatible with the mean field value $1/\delta_h =
1/2$~\cite{Marro99}.

Results in Fig.~\ref{fig:response} are typical, similar curves are
obtained for any $m>1$. The performance of these scale-free networks
in enhancing the dynamic range is poor: while the dynamic range of
isolated excitable elements is $\Delta(\sigma=0)\simeq 16.7$~dB, the
network (optimal) performance at criticality is only $\Delta(\sigma_c)
\simeq 20.8$~dB, an enhancement of less than a decade. This is
slightly worse than the enhancement produced by random networks with
equivalent size and average connectivity, as can be seen in the curves
$\Delta(\sigma)$ of Fig.~\ref{fig:deltasigma}.

The case $m=1$, however, is particularly interesting. Notice that in
this situation the network is still scale-free, but does not comprise
any loop in its structure and consequently has a tree-like
pattern. This condition, together with the deterministic nature of
each excitable node after excitation, prevents the phase transition to
self-sustained activity from occurring~\cite{lewis00}, a fact that has
also been observed in one-dimensional excitable
networks~\cite{Furtado06}. In these conditions the only transition
occurs at $\sigma=\sigma_{max} = K$ ($=2m$ for scale-free networks),
whereby propagation of excitable waves becomes deterministic
(ballistic). Therefore low-stimulus amplification increases steadily
with $\sigma$, but in the absence of self-sustained activity
($F_0=0$). This allows the dynamic range to increase monotonically
with $\sigma$, reaching values near 50~dB, which is the largest value
obtained so far in excitable network models.

\subsection{\label{ld}Loop-diluted model}

As we observe a remarkable difference between the dynamic range of
scale-free networks with $m=1$ and other values of $m$, and the former
has a typical feature (non-existence of loops) which is not present in
$m>1$ topologies, we are interested in investigating the role of loops
in the response functions of the networks.  For this purpose, we
propose a variant of the BA model which is referred to as loop-diluted
model. In the model each new node is added according to the usual
preferential attachment rule, but can have $m=1$ or $m=2$ links
according to the probability distribution
\begin{equation}
P(m)=(1-p)\delta_{m,1}+p\delta_{m,2}\; ,
\end{equation}
where $p$ is the probability of having two edges. So $p$ adjusts the
amount of loops in the network and the case $p=0$ recovers the
structure with no loops. The mean degree is now $K = 2\left< m\right>
= 2(1+p)$.

\begin{figure}[t]
\centerline{\includegraphics[width=0.95\columnwidth]{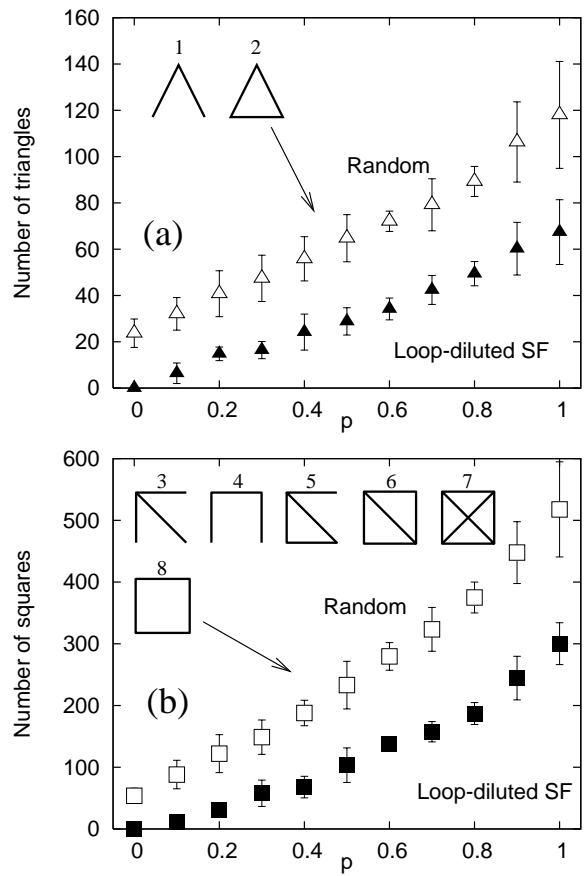}}
\caption{\label{fig:motif} Occurrences of 3- and 4-site patterns in
  the loop-diluted scale-free networks (filled symbols) and their
  randomized counterparts (open symbols). (a) Number of triangles
  versus $p$ and (b) Number of squares versus $p$. Points (bars)
  represent the mean (standard deviation) over 5 realizations for
  $N=1000$. Patterns are sequentially numbered (1-8) for further
  reference in the text.}
\end{figure}

Notice that, since all sites belong to a single giant component, the
average number of loops created by each newly added site is bounded
from below by $p$. Each new two-edged node can give rise to loops of
any size, but the relationship between parameter $p$ and the number of
loops becomes already apparent in a simple 3-site motif
analysis~\cite{Milo02}. Figure~\ref{fig:motif}(a) shows the mean
number of triangles as a function of $p$ (calculated by the free
software available at {\tt www.weizman.ac.il/mcb/UriAlon}). As
expected, this is a monotonically increasing function, which
nevertheless stays well below $Np$, hinting that most loops comprise
more than three sites. The same qualitative scenario is observed when
we plot the number of squares (Fig.~\ref{fig:motif}(b)), which are
more abundant than triangles. In both cases, we notice that for
equally sized randomized graphs (which preserve the degrees of every
node~\cite{Milo02}), the numbers of triangles and squares are
considerably larger. This means that triangles and squares are
actually {\em anti-motifs\/} in the loop-diluted
model~\cite{Milo04}. In fact, this is true for all patterns that
contain loops (numbered 2, 5, 6 and 8 in Fig.~\ref{fig:motif}), except
for pattern 7, which cannot occur according to the growth rules of the
model. The occurrences of patterns 1, 3 and 4 in the loop-diluted
model and in the randomized networks are statistically
indistinguishable (therefore they are neither motifs nor anti-motifs
-- data not shown).

\begin{figure}[t]
\centerline{\includegraphics[width=0.95\columnwidth]{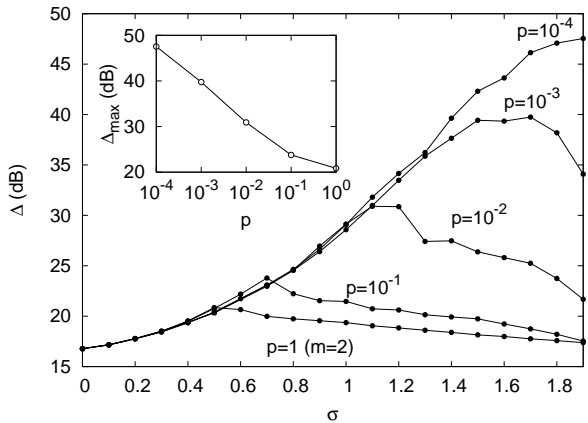}}
\caption{\label{fig:figp} Dynamic range $\Delta$ versus branching
parameter $\sigma$ for the loop-diluted model and different values of
the probability $p$ that a newly added node has two incoming
edges. Inset: maximum value of the dynamic range as a function of
$p$.}
\end{figure}

Figure \ref{fig:figp} displays the dynamic range $\Delta$ as a
function of the coupling $\sigma$ for some values of $p$. From the
figure, we clearly notice that the insertion of loops by increasing
the probability $p$ has a striking effect on the dynamic range. This
effect is already mensurable for values of $p$ such that $pN \sim
1$. The peak value of the dynamic range $\Delta_{max}(p)\equiv
\max_{\sigma}\Delta(\sigma;p)=\Delta(\sigma_c(p);p)$ seems to decrease
logarithmically with $p$.

\section{\label{conclusions}Concluding remarks}

Recently, some investigations have addressed the enlargement in
average activity of excitable elements by coupling these entities and
so giving form to a new larger and more sensitive unit. Although this
collective phenomenon has been widely accepted, little is known about
the way the arrangement of connections among the excitable elements
acts physically on the system dynamics. The creation of more robust
and functional units from smaller units (whose pattern of interactions
is a determining aspect) is of course not exclusive of
Neuroscience. For instance, the hypercycle, a catalytic feedback
network whereby each element helps the replication of the next one in
a regulatory cycle closing on itself, has been pondered as an
alternative resolution for the information crisis in prebiotic
evolution~\cite{Eigen78,Campos00}.  We believe that all the recent
contributions on this issue have a bearing on a more general context,
that is, the understanding of the interplay between system dynamics
and the underlying interaction network topologies. We hope that our
contribution gives a small step in this direction, when we corroborate
that the amount of loops in network structure could be a key
topological feature. 

We have presented simulation results for the transfer function of
excitable scale-free networks. The behavior of the dynamic range
$\Delta$ as a function of the coupling $\sigma$ shows the general
behavior predicted in Ref.~\cite{Kinouchi06a}: in the subcritical
regime ($\sigma<\sigma_c$) $\Delta(\sigma)$ increases, while in the
supercritical regime ($\sigma>\sigma_c$) $\Delta(\sigma)$
decreases. The maximum value is obtained at criticality, but for
scale-free networks with $m>1$ this result is even smaller than that for
a random graph. 

For $m=1$ the phase transition to self-sustained activity disappears,
and the dynamic range increases steadily, reaching its maximum value
when excitable waves become deterministic. This suggests that the
presence of loops in the network could be a relevant feature in
determining the dynamic range of its transfer function. We have
introduced a simple extension to the BA scale-free model which allows
one to interpolate between $m=1$ and $m=2$, showing that dynamic range
decreases as the density of loops increases. This reinforces the need
to study other topologies with tree structure, which abound in
biological structures.

It remains at present unclear whether the response exponent for $m>1$
is indeed compatible with the mean field universality class (even
though this seems to be supported by recent simulation results in
Ref.~\cite{Wu07}, which independently addressed a similar
problem). Also, for $m=1$ at maximum coupling, the response function
seems to be governed by a power law with a much smaller exponent (see
inset of Fig.~\ref{fig:deltasigma}) which might not belong to the
directed percolation universality class. We believe that a detailed
study of the critical exponents of excitable scale-free networks is
still lacking and should be dealt with in the future.


MC and PRAC are supported by Conselho Nacional de Desenvolvimento
Cient\'{\i}fico e Tecnol\'ogico (CNPq), FACEPE and special program
PRONEX.

\bibliography{copelli}

\begin{thebibliography}{35}
\expandafter\ifx\csname natexlab\endcsname\relax\def\natexlab#1{#1}\fi
\expandafter\ifx\csname bibnamefont\endcsname\relax
  \def\bibnamefont#1{#1}\fi
\expandafter\ifx\csname bibfnamefont\endcsname\relax
  \def\bibfnamefont#1{#1}\fi
\expandafter\ifx\csname citenamefont\endcsname\relax
  \def\citenamefont#1{#1}\fi
\expandafter\ifx\csname url\endcsname\relax
  \def\url#1{\texttt{#1}}\fi
\expandafter\ifx\csname urlprefix\endcsname\relax\def\urlprefix{URL }\fi
\providecommand{\bibinfo}[2]{#2}
\providecommand{\eprint}[2][]{\url{#2}}

\bibitem[{\citenamefont{Lewis and Rinzel}(2000)}]{lewis00}
\bibinfo{author}{\bibfnamefont{T.~J.} \bibnamefont{Lewis}} \bibnamefont{and}
  \bibinfo{author}{\bibfnamefont{J.}~\bibnamefont{Rinzel}},
  \bibinfo{journal}{Network: Comput. Neural Syst.}
  \textbf{\bibinfo{volume}{11}}, \bibinfo{pages}{299} (\bibinfo{year}{2000}).

\bibitem[{\citenamefont{Copelli et~al.}(2002)\citenamefont{Copelli, Roque,
  Oliveira, and Kinouchi}}]{Copelli02}
\bibinfo{author}{\bibfnamefont{M.}~\bibnamefont{Copelli}},
  \bibinfo{author}{\bibfnamefont{A.~C.} \bibnamefont{Roque}},
  \bibinfo{author}{\bibfnamefont{R.~F.} \bibnamefont{Oliveira}},
  \bibnamefont{and} \bibinfo{author}{\bibfnamefont{O.}~\bibnamefont{Kinouchi}},
  \bibinfo{journal}{Phys. Rev. E} \textbf{\bibinfo{volume}{65}},
  \bibinfo{pages}{060901} (\bibinfo{year}{2002}).

\bibitem[{\citenamefont{Copelli et~al.}(2005)\citenamefont{Copelli, Oliveira,
  Roque, and Kinouchi}}]{Copelli05a}
\bibinfo{author}{\bibfnamefont{M.}~\bibnamefont{Copelli}},
  \bibinfo{author}{\bibfnamefont{R.~F.} \bibnamefont{Oliveira}},
  \bibinfo{author}{\bibfnamefont{A.~C.} \bibnamefont{Roque}}, \bibnamefont{and}
  \bibinfo{author}{\bibfnamefont{O.}~\bibnamefont{Kinouchi}},
  \bibinfo{journal}{Neurocomputing} \textbf{\bibinfo{volume}{65-66}},
  \bibinfo{pages}{691} (\bibinfo{year}{2005}).

\bibitem[{\citenamefont{Copelli and Kinouchi}(2005)}]{Copelli05b}
\bibinfo{author}{\bibfnamefont{M.}~\bibnamefont{Copelli}} \bibnamefont{and}
  \bibinfo{author}{\bibfnamefont{O.}~\bibnamefont{Kinouchi}},
  \bibinfo{journal}{Physica A} \textbf{\bibinfo{volume}{349}},
  \bibinfo{pages}{431} (\bibinfo{year}{2005}).

\bibitem[{\citenamefont{Haldeman and Beggs}(2005)}]{Haldeman05}
\bibinfo{author}{\bibfnamefont{C.}~\bibnamefont{Haldeman}} \bibnamefont{and}
  \bibinfo{author}{\bibfnamefont{J.~M.} \bibnamefont{Beggs}},
  \bibinfo{journal}{Phys. Rev. Lett.} \textbf{\bibinfo{volume}{94}},
  \bibinfo{pages}{058101} (\bibinfo{year}{2005}).

\bibitem[{\citenamefont{Furtado and Copelli}(2006)}]{Furtado06}
\bibinfo{author}{\bibfnamefont{L.~S.} \bibnamefont{Furtado}} \bibnamefont{and}
  \bibinfo{author}{\bibfnamefont{M.}~\bibnamefont{Copelli}},
  \bibinfo{journal}{Phys. Rev. E} \textbf{\bibinfo{volume}{73}},
  \bibinfo{pages}{011907} (\bibinfo{year}{2006}).

\bibitem[{\citenamefont{Kinouchi and Copelli}(2006)}]{Kinouchi06a}
\bibinfo{author}{\bibfnamefont{O.}~\bibnamefont{Kinouchi}} \bibnamefont{and}
  \bibinfo{author}{\bibfnamefont{M.}~\bibnamefont{Copelli}},
  \bibinfo{journal}{Nat. Phys.} \textbf{\bibinfo{volume}{2}},
  \bibinfo{pages}{348} (\bibinfo{year}{2006}).

\bibitem[{\citenamefont{Doiron et~al.}(2006)\citenamefont{Doiron, Rinzel, and
  Reyes}}]{Doiron06}
\bibinfo{author}{\bibfnamefont{B.}~\bibnamefont{Doiron}},
  \bibinfo{author}{\bibfnamefont{J.}~\bibnamefont{Rinzel}}, \bibnamefont{and}
  \bibinfo{author}{\bibfnamefont{A.}~\bibnamefont{Reyes}},
  \bibinfo{journal}{Phys. Rev. E} \textbf{\bibinfo{volume}{74}},
  \bibinfo{pages}{030903} (\bibinfo{year}{2006}).

\bibitem[{\citenamefont{Watts and Strogatz}(1998)}]{Watts98}
\bibinfo{author}{\bibfnamefont{D.~J.} \bibnamefont{Watts}} \bibnamefont{and}
  \bibinfo{author}{\bibfnamefont{S.~H.} \bibnamefont{Strogatz}},
  \bibinfo{journal}{Nature} \textbf{\bibinfo{volume}{393}},
  \bibinfo{pages}{440} (\bibinfo{year}{1998}).

\bibitem[{\citenamefont{Amaral et~al.}(2000)\citenamefont{Amaral, Scala,
  Barth\'el\'emy, and Stanley}}]{Amaral00}
\bibinfo{author}{\bibfnamefont{L.~A.~N.} \bibnamefont{Amaral}},
  \bibinfo{author}{\bibfnamefont{A.}~\bibnamefont{Scala}},
  \bibinfo{author}{\bibfnamefont{M.}~\bibnamefont{Barth\'el\'emy}},
  \bibnamefont{and} \bibinfo{author}{\bibfnamefont{H.~E.}
  \bibnamefont{Stanley}}, \bibinfo{journal}{PNAS}
  \textbf{\bibinfo{volume}{97}}, \bibinfo{pages}{11149} (\bibinfo{year}{2000}).

\bibitem[{\citenamefont{Sakata et~al.}(2005)\citenamefont{Sakata, Komatsu, and
  Yamamori}}]{Sakata05}
\bibinfo{author}{\bibfnamefont{S.}~\bibnamefont{Sakata}},
  \bibinfo{author}{\bibfnamefont{Y.}~\bibnamefont{Komatsu}}, \bibnamefont{and}
  \bibinfo{author}{\bibfnamefont{T.}~\bibnamefont{Yamamori}},
  \bibinfo{journal}{Neurosci. Res.} \textbf{\bibinfo{volume}{51}},
  \bibinfo{pages}{309} (\bibinfo{year}{2005}).

\bibitem[{\citenamefont{Beggs and Plenz}(2003)}]{Beggs03}
\bibinfo{author}{\bibfnamefont{J.~M.} \bibnamefont{Beggs}} \bibnamefont{and}
  \bibinfo{author}{\bibfnamefont{D.}~\bibnamefont{Plenz}}, \bibinfo{journal}{J.
  Neurosci.} \textbf{\bibinfo{volume}{23}}, \bibinfo{pages}{11167}
  (\bibinfo{year}{2003}).

\bibitem[{\citenamefont{Egu\'{\i}luz et~al.}(2005)\citenamefont{Egu\'{\i}luz,
  Chialvo, Cecchi, Baliki, and Apkarian}}]{Eguiluz05}
\bibinfo{author}{\bibfnamefont{V.~M.} \bibnamefont{Egu\'{\i}luz}},
  \bibinfo{author}{\bibfnamefont{D.~R.} \bibnamefont{Chialvo}},
  \bibinfo{author}{\bibfnamefont{G.~A.} \bibnamefont{Cecchi}},
  \bibinfo{author}{\bibfnamefont{M.}~\bibnamefont{Baliki}}, \bibnamefont{and}
  \bibinfo{author}{\bibfnamefont{A.~V.} \bibnamefont{Apkarian}},
  \bibinfo{journal}{Phys. Rev. Lett.} \textbf{\bibinfo{volume}{94}},
  \bibinfo{pages}{018102} (\bibinfo{year}{2005}).

\bibitem[{\citenamefont{Chialvo}(2004)}]{Chialvo04}
\bibinfo{author}{\bibfnamefont{D.~R.} \bibnamefont{Chialvo}},
  \bibinfo{journal}{Physica A} \textbf{\bibinfo{volume}{340}},
  \bibinfo{pages}{756} (\bibinfo{year}{2004}).

\bibitem[{\citenamefont{Sporns et~al.}(2004)\citenamefont{Sporns, Chialvo,
  Kaiser, and Hilgetag}}]{Sporns04a}
\bibinfo{author}{\bibfnamefont{O.}~\bibnamefont{Sporns}},
  \bibinfo{author}{\bibfnamefont{D.~R.} \bibnamefont{Chialvo}},
  \bibinfo{author}{\bibfnamefont{M.}~\bibnamefont{Kaiser}}, \bibnamefont{and}
  \bibinfo{author}{\bibfnamefont{C.~C.} \bibnamefont{Hilgetag}},
  \bibinfo{journal}{Trends Cog. Sci.} \textbf{\bibinfo{volume}{8}},
  \bibinfo{pages}{418} (\bibinfo{year}{2004}).

\bibitem[{\citenamefont{Stevens}(1975)}]{Stevens}
\bibinfo{author}{\bibfnamefont{S.~S.} \bibnamefont{Stevens}},
  \emph{\bibinfo{title}{Psychophysics: Introduction to its Perceptual, Neural
  and Social Prospects}} (\bibinfo{publisher}{Wiley, New York},
  \bibinfo{year}{1975}).

\bibitem[{\citenamefont{Firestein et~al.}(1993)\citenamefont{Firestein, Picco,
  and Menini}}]{Firestein93}
\bibinfo{author}{\bibfnamefont{S.}~\bibnamefont{Firestein}},
  \bibinfo{author}{\bibfnamefont{C.}~\bibnamefont{Picco}}, \bibnamefont{and}
  \bibinfo{author}{\bibfnamefont{A.}~\bibnamefont{Menini}},
  \bibinfo{journal}{J. Physiol.} \textbf{\bibinfo{volume}{468}},
  \bibinfo{pages}{1} (\bibinfo{year}{1993}).

\bibitem[{\citenamefont{Rospars et~al.}(2000)\citenamefont{Rospars, L\'ansk\'y,
  Duchamp-Viret, and Duchamp}}]{Rospars00}
\bibinfo{author}{\bibfnamefont{J.-P.} \bibnamefont{Rospars}},
  \bibinfo{author}{\bibfnamefont{P.}~\bibnamefont{L\'ansk\'y}},
  \bibinfo{author}{\bibfnamefont{P.}~\bibnamefont{Duchamp-Viret}},
  \bibnamefont{and} \bibinfo{author}{\bibfnamefont{A.}~\bibnamefont{Duchamp}},
  \bibinfo{journal}{BioSystems} \textbf{\bibinfo{volume}{58}},
  \bibinfo{pages}{133} (\bibinfo{year}{2000}).

\bibitem[{\citenamefont{Rospars et~al.}(2003)\citenamefont{Rospars, L\'ansk\'y,
  Duchamp-Viret, and Duchamp}}]{Rospars03}
\bibinfo{author}{\bibfnamefont{J.-P.} \bibnamefont{Rospars}},
  \bibinfo{author}{\bibfnamefont{P.}~\bibnamefont{L\'ansk\'y}},
  \bibinfo{author}{\bibfnamefont{P.}~\bibnamefont{Duchamp-Viret}},
  \bibnamefont{and} \bibinfo{author}{\bibfnamefont{A.}~\bibnamefont{Duchamp}},
  \bibinfo{journal}{Eur. J. Neurosci.} \textbf{\bibinfo{volume}{18}},
  \bibinfo{pages}{1135} (\bibinfo{year}{2003}).

\bibitem[{\citenamefont{Deans et~al.}(2002)\citenamefont{Deans, Volgyi,
  Goodenough, Bloomfield, and Paul}}]{Deans02}
\bibinfo{author}{\bibfnamefont{M.~R.} \bibnamefont{Deans}},
  \bibinfo{author}{\bibfnamefont{B.}~\bibnamefont{Volgyi}},
  \bibinfo{author}{\bibfnamefont{D.~A.} \bibnamefont{Goodenough}},
  \bibinfo{author}{\bibfnamefont{S.~A.} \bibnamefont{Bloomfield}},
  \bibnamefont{and} \bibinfo{author}{\bibfnamefont{D.~L.} \bibnamefont{Paul}},
  \bibinfo{journal}{Neuron} \textbf{\bibinfo{volume}{36}}, \bibinfo{pages}{703}
  (\bibinfo{year}{2002}).

\bibitem[{\citenamefont{Greenberg and Hastings}(1978)}]{Greenberg78}
\bibinfo{author}{\bibfnamefont{J.~M.} \bibnamefont{Greenberg}}
  \bibnamefont{and} \bibinfo{author}{\bibfnamefont{S.~P.}
  \bibnamefont{Hastings}}, \bibinfo{journal}{{SIAM} J. Appl. Math.}
  \textbf{\bibinfo{volume}{34}}, \bibinfo{pages}{515} (\bibinfo{year}{1978}).

\bibitem[{\citenamefont{Marro and Dickman}(1999)}]{Marro99}
\bibinfo{author}{\bibfnamefont{J.}~\bibnamefont{Marro}} \bibnamefont{and}
  \bibinfo{author}{\bibfnamefont{R.}~\bibnamefont{Dickman}},
  \emph{\bibinfo{title}{Nonequilibrium Phase Transition in Lattice Models}}
  (\bibinfo{publisher}{Cambridge University Press},
  \bibinfo{address}{Cambridge}, \bibinfo{year}{1999}).

\bibitem[{\citenamefont{Barab\'asi and Albert}(1999)}]{Barabasi99Sci}
\bibinfo{author}{\bibfnamefont{A.-L.} \bibnamefont{Barab\'asi}}
  \bibnamefont{and} \bibinfo{author}{\bibfnamefont{R.}~\bibnamefont{Albert}},
  \bibinfo{journal}{Science} \textbf{\bibinfo{volume}{286}},
  \bibinfo{pages}{509} (\bibinfo{year}{1999}).

\bibitem[{\citenamefont{Newman}(2001)}]{Newman01e}
\bibinfo{author}{\bibfnamefont{M.~E.~J.} \bibnamefont{Newman}},
  \bibinfo{journal}{Proc. Natl. Acad. Sci.} \textbf{\bibinfo{volume}{98}},
  \bibinfo{pages}{404} (\bibinfo{year}{2001}).

\bibitem[{\citenamefont{Moreno et~al.}(2002)\citenamefont{Moreno,
  Pastor-Satorras, and Vespignani}}]{Moreno02}
\bibinfo{author}{\bibfnamefont{Y.}~\bibnamefont{Moreno}},
  \bibinfo{author}{\bibfnamefont{R.}~\bibnamefont{Pastor-Satorras}},
  \bibnamefont{and}
  \bibinfo{author}{\bibfnamefont{A.}~\bibnamefont{Vespignani}},
  \bibinfo{journal}{Eur. Phys. J. B} \textbf{\bibinfo{volume}{26}},
  \bibinfo{pages}{521} (\bibinfo{year}{2002}).

\bibitem[{\citenamefont{Vuorinen et~al.}(2004)\citenamefont{Vuorinen,
  Peltom\"aki, Rost, and Alava}}]{Vuorinen04}
\bibinfo{author}{\bibfnamefont{V.}~\bibnamefont{Vuorinen}},
  \bibinfo{author}{\bibfnamefont{M.}~\bibnamefont{Peltom\"aki}},
  \bibinfo{author}{\bibfnamefont{M.}~\bibnamefont{Rost}}, \bibnamefont{and}
  \bibinfo{author}{\bibfnamefont{M.~J.} \bibnamefont{Alava}},
  \bibinfo{journal}{Eur. Phys. J. B} \textbf{\bibinfo{volume}{38}},
  \bibinfo{pages}{261} (\bibinfo{year}{2004}).

\bibitem[{\citenamefont{Jeong et~al.}(2000)\citenamefont{Jeong, Tombor, Albert,
  Oltvai, and Barab\'asi}}]{Jeong00metabolic}
\bibinfo{author}{\bibfnamefont{H.}~\bibnamefont{Jeong}},
  \bibinfo{author}{\bibfnamefont{B.}~\bibnamefont{Tombor}},
  \bibinfo{author}{\bibfnamefont{R.}~\bibnamefont{Albert}},
  \bibinfo{author}{\bibfnamefont{Z.~N.} \bibnamefont{Oltvai}},
  \bibnamefont{and} \bibinfo{author}{\bibfnamefont{A.-L.}
  \bibnamefont{Barab\'asi}}, \bibinfo{journal}{Nature}
  \textbf{\bibinfo{volume}{407}}, \bibinfo{pages}{651} (\bibinfo{year}{2000}).

\bibitem[{\citenamefont{Jeong et~al.}(2001)\citenamefont{Jeong, Mason,
  Barab\'asi, and Oltvai}}]{Jeong01lethality}
\bibinfo{author}{\bibfnamefont{H.}~\bibnamefont{Jeong}},
  \bibinfo{author}{\bibfnamefont{S.~P.} \bibnamefont{Mason}},
  \bibinfo{author}{\bibfnamefont{A.-L.} \bibnamefont{Barab\'asi}},
  \bibnamefont{and} \bibinfo{author}{\bibfnamefont{Z.~N.}
  \bibnamefont{Oltvai}}, \bibinfo{journal}{Nature}
  \textbf{\bibinfo{volume}{411}}, \bibinfo{pages}{41} (\bibinfo{year}{2001}).

\bibitem[{\citenamefont{Albert and Barab\'asi}(2002)}]{Albert02}
\bibinfo{author}{\bibfnamefont{R.}~\bibnamefont{Albert}} \bibnamefont{and}
  \bibinfo{author}{\bibfnamefont{A.-L.} \bibnamefont{Barab\'asi}},
  \bibinfo{journal}{Rev. Mod. Phys.} \textbf{\bibinfo{volume}{74}},
  \bibinfo{pages}{47} (\bibinfo{year}{2002}).

\bibitem[{\citenamefont{Castellano and Pastor-Satorras}(2006)}]{Castellano06}
\bibinfo{author}{\bibfnamefont{C.}~\bibnamefont{Castellano}} \bibnamefont{and}
  \bibinfo{author}{\bibfnamefont{R.}~\bibnamefont{Pastor-Satorras}},
  \bibinfo{journal}{Phys. Rev. Lett.} \textbf{\bibinfo{volume}{96}},
  \bibinfo{pages}{038701} (\bibinfo{year}{2006}).

\bibitem[{\citenamefont{Milo et~al.}(2002)\citenamefont{Milo, Shen-Orr,
  Itzkovitz, Kashtan, Chklovskii, and Alon}}]{Milo02}
\bibinfo{author}{\bibfnamefont{R.}~\bibnamefont{Milo}},
  \bibinfo{author}{\bibfnamefont{S.}~\bibnamefont{Shen-Orr}},
  \bibinfo{author}{\bibfnamefont{S.}~\bibnamefont{Itzkovitz}},
  \bibinfo{author}{\bibfnamefont{N.}~\bibnamefont{Kashtan}},
  \bibinfo{author}{\bibfnamefont{D.}~\bibnamefont{Chklovskii}},
  \bibnamefont{and} \bibinfo{author}{\bibfnamefont{U.}~\bibnamefont{Alon}},
  \bibinfo{journal}{Science} \textbf{\bibinfo{volume}{298}},
  \bibinfo{pages}{824} (\bibinfo{year}{2002}).

\bibitem[{\citenamefont{Milo et~al.}(2004)\citenamefont{Milo, Itzkovitz,
  Kashtan, Levitt, Shen-Orr, Ayzenshtat, Sheffer, and Alon}}]{Milo04}
\bibinfo{author}{\bibfnamefont{R.}~\bibnamefont{Milo}},
  \bibinfo{author}{\bibfnamefont{S.}~\bibnamefont{Itzkovitz}},
  \bibinfo{author}{\bibfnamefont{N.}~\bibnamefont{Kashtan}},
  \bibinfo{author}{\bibfnamefont{R.}~\bibnamefont{Levitt}},
  \bibinfo{author}{\bibfnamefont{S.}~\bibnamefont{Shen-Orr}},
  \bibinfo{author}{\bibfnamefont{I.}~\bibnamefont{Ayzenshtat}},
  \bibinfo{author}{\bibfnamefont{M.}~\bibnamefont{Sheffer}}, \bibnamefont{and}
  \bibinfo{author}{\bibfnamefont{U.}~\bibnamefont{Alon}},
  \bibinfo{journal}{Science} \textbf{\bibinfo{volume}{303}},
  \bibinfo{pages}{1538} (\bibinfo{year}{2004}).

\bibitem[{\citenamefont{Eigen and P.}(1978)}]{Eigen78}
\bibinfo{author}{\bibfnamefont{M.}~\bibnamefont{Eigen}} \bibnamefont{and}
  \bibinfo{author}{\bibfnamefont{S.}~\bibnamefont{P.}},
  \bibinfo{journal}{Naturwissenschaften} \textbf{\bibinfo{volume}{65}},
  \bibinfo{pages}{7} (\bibinfo{year}{1978}).

\bibitem[{\citenamefont{Campos et~al.}(2000)\citenamefont{Campos, Fontanari,
  and Stadler}}]{Campos00}
\bibinfo{author}{\bibfnamefont{P.~R.~A.} \bibnamefont{Campos}},
  \bibinfo{author}{\bibfnamefont{J.~F.} \bibnamefont{Fontanari}},
  \bibnamefont{and} \bibinfo{author}{\bibfnamefont{P.~F.}
  \bibnamefont{Stadler}}, \bibinfo{journal}{Phys. Rev. E}
  \textbf{\bibinfo{volume}{61}}, \bibinfo{pages}{2996} (\bibinfo{year}{2000}).

\bibitem[{\citenamefont{Wu et~al.}(2007)\citenamefont{Wu, Xu, and Wang}}]{Wu07}
\bibinfo{author}{\bibfnamefont{A.-C.} \bibnamefont{Wu}},
  \bibinfo{author}{\bibfnamefont{X.-J.} \bibnamefont{Xu}}, \bibnamefont{and}
  \bibinfo{author}{\bibfnamefont{Y.-H.} \bibnamefont{Wang}},
  \bibinfo{journal}{Phys. Rev. E} \textbf{\bibinfo{volume}{75}},
  \bibinfo{pages}{032901} (\bibinfo{year}{2007}).

\end{thebibliography}

\end{document}